\documentclass[pdflatex,sn-nature]{sn-jnl}

\usepackage{graphicx}%
\usepackage{multirow}%
\usepackage{amsmath,amssymb,amsfonts}%
\usepackage{amsthm}%
\usepackage{mathrsfs}%
\usepackage[title]{appendix}%
\usepackage{xcolor}%
\usepackage{textcomp}%
\usepackage{manyfoot}%
\usepackage{booktabs}%
\usepackage{algorithm}%
\usepackage{algorithmicx}%
\usepackage{algpseudocode}%
\usepackage{listings}%
\usepackage{setspace}
\usepackage{float}
\usepackage{placeins}
\usepackage{geometry}
\theoremstyle{thmstyleone}%
\theoremstyle{thmstyletwo}%

\theoremstyle{thmstylethree}%

\begin{document}

\title[Article Title]{Harnessing disorder to decouple extension and shear in kirigami metamaterials}

\author[1,2]{\fnm{Haomin} \sur{Yu}}
\equalcont{These authors contributed equally to this work.}

\author*[1,2,3]{\fnm{Hanxun} \sur{Jin}}\email{jinhx@ucmail.uc.edu}
\equalcont{These authors contributed equally to this work.}

\author[1,2]{\fnm{Mingxuan} \sur{Bi}}
\equalcont{These authors contributed equally to this work.}

\author[4]{\fnm{Mohammad} \sur{Jafari}}

\author[1,2]{\fnm{Feng Helen} \sur{Long}}

\author[1,5]{\fnm{Michael J.} \sur{Greenberg}}

\author*[4]{\fnm{Farid} \sur{Alisafaei}}\email{farid.alisafaei@njit.edu}

\author*[1,2]{\fnm{Guy M.} \sur{Genin}}\email{genin@wustl.edu}

\affil[1]{\orgdiv{NSF Science and Technology Center for Engineering MechanoBiology}, \orgname{Washington University in St. Louis}, \orgaddress{\city{St. Louis}, \state{Missouri}, \postcode{63130}, \country{USA}}}

\affil[2]{\orgdiv{Department of Mechanical Engineering \& Materials Science}, \orgname{Washington University in St. Louis}, \orgaddress{\city{St. Louis}, \state{Missouri}, \postcode{63130}, \country{USA}}}

\affil[3]{\orgdiv{Department of Mechanical and Materials Engineering}, \orgname{University of Cincinnati}, \orgaddress{\city{Cincinnati}, \state{Ohio}, \postcode{45221}, \country{USA}}}

\affil[4]{\orgdiv{Department of Mechanical Engineering}, \orgname{New Jersey Institute of Technology}, \orgaddress{\city{Newark}, \state{New Jersey}, \postcode{07102}, \country{USA}}}

\affil[5]{\orgdiv{Department of Biochemistry and Molecular Biophysics}, \orgname{Washington University School of Medicine}, \orgaddress{\city{St. Louis}, \state{Missouri}, \postcode{63110}, \country{USA}}}

\abstract{Kirigami turns stiff sheets into compliant, shape-morphing structures, but its reliance on periodic cut patterns comes at a cost: correlated panel rotations couple extension to shear, so stretching one axis drives a parasitic shear that cannot be suppressed, and also confine anisotropic stiffness to a narrow, discrete set of responses that cannot be tuned independently. Biological tissues overcome an analogous constraint through controlled disorder, such as graded fiber orientations in skin and hierarchical anisotropy in myocardium, achieving direction-dependent mechanics unavailable to regular architectures. Here, we show that engineered disorder is a design degree of freedom for kirigami, with stochastic kirigami accessing a continuous and far broader region of mechanical response than periodic patterns. This includes programmable anisotropy with near-complete elimination of extension-shear coupling. Because disordered patterns lack a simple parameterization, we navigate this design space with a geometry-aware graph neural network (GNN) that maps cut topology to the full nonlinear, bidirectional stress-strain response, coupled to a genetic algorithm that inverse-designs patterns reproducing target responses along two perpendicular axes. The GNN trains an order of magnitude faster and more accurately than image-based models. Fabricated elastomer samples reproduce the predicted nonlinear, anisotropic responses, closing the loop from design to physical component. By turning disorder into a variable to control directional stiffness, this work develops architected materials that stretch without parasitic shear, from soft actuators to tissue-interfacing devices matched to the anisotropy of living tissue.}


\maketitle

\clearpage

\section{Introduction}\label{sec1}

The patterning of a thin sheet with kirigami arrays of cuts has become one of the most versatile routes to mechanical metamaterials, materials whose properties arise from architecture rather than composition \cite{bertoldi2017flexible, jin2024mechanical, jiao2023mechanical, surjadi2019mechanical, jin2026situ}. 
By transforming stiff, inextensible substrates into compliant, stretchable, and shape-morphing structures \cite{jin2025characterization}, strategically placed incisions \cite{shyu2015kirigami, blees2015graphene} have enabled advances that depend on large, reversible deformation: stretchable electronics \cite{zhang2015mechanically}, conformal bioelectronics \cite{meng2022kirigami}, and tissue-interfacing devices \cite{brooks2022kirigami} that must move with the soft, anisotropic tissues they contact. 
However, almost all of these designs share a single structural assumption, namely that the cuts repeat in a periodic or symmetric pattern. 
This regularity simplifies fabrication and analysis, but, as we show here, it is not a neutral choice: periodicity imposes a kinematic constraint that fundamentally limits the mechanical behavior kirigami can achieve.

The source of this fundamental limit on the properties of kirigami materials is the repeating motif itself.
In a periodic kirigami sheet, identical unit cells rotate in a correlated fashion under load, so that stretching along one axis drives a coordinated, biased shear and locks the in-plane anisotropy onto a narrow, discrete envelope of stiffness ratios.
These properties can be traded against one another but not set independently \cite{rafsanjani2017buckling}. 
Biological tissues meet the same mechanical demands, but have evolved to escape this constraint, because they are built from architectures defined not by periodicity but by controlled disorder \cite{genin2017unification, golman2021toughening, zaiser2023disordered}.
Heterogeneous collagen fiber orientations give skin its direction-dependent compliance, the basis of Langer lines \cite{annaidh2012characterization}; hierarchically organized cardiomyocyte bundles and perimysial collagen sheets endow myocardium with its orthotropic response \cite{tueni2023structural}; and disordered fibrillar networks achieve coordination-dependent strain stiffening \cite{jansen2018role}. 
In each, heterogeneity and aperiodicity, typically avoided in engineering materials, become functional design elements that deliver mechanical performance unreachable by regular architectures.

We therefore hypothesized that disorder could play the same role in engineered kirigami, if it could be controlled and deliberately designed. 
Recent work supports the premise. 
Chen et al. established the topological foundations of random cut networks, deriving percolation thresholds and showing that stochastic patterns furnish degrees of freedom unavailable to deterministic ones \cite{chen2020deterministic}, while Chaudhary et al. characterized the mechanics of randomly cut sheets across deformation regimes \cite{chaudhary2023geometric}. 
More broadly, disorder has been shown to enhance fracture toughness \cite{fulco2025disorder}, relax the strength-toughness trade-off \cite{choukir2025disorder}, and generate auxetic behavior through selective network pruning \cite{reid2018auxetic}. 
The obstacle is therefore not whether disorder is useful but whether it can be harnessed: the design space of aperiodic cut patterns is combinatorially vast and, unlike the low-dimensional parameterizations that make periodic lattices tractable, admits no simple description by which it could be searched systematically.

We explored a data-driven modeling approach to enable this.
Machine-learning surrogates can learn structure-property relationships from simulation and evaluate candidate designs fast enough to operate inside an optimization loop \cite{bastek2023inverse, ha2023rapid, jin2023recent, zheng2023deep, jin2025characterization}. 
For architectures whose behavior is governed by connectivity rather than absolute position, the representation is critical: convolutional neural networks operate on pixelated images that blur topological structure, whereas graph neural networks (GNNs) encode cuts and their interactions directly as nodes and edges \cite{reiser2022graph, khemani2024review}. 
This advantage is well-documented for polycrystalline microstructures \cite{dai2021graph} and truss lattices \cite{zheng2023unifying, frey2025multi, maurizi2025designing}, and GNN surrogates have been coupled with genetic, Bayesian, and differentiable optimizers to inverse-design metamaterials with prescribed nonlinear responses \cite{maurizi2025designing, dold2023differentiable, bastek2023inverse}. 
However, no such framework has been built for kirigami, despite the natural correspondence between cut patterns and graphs.
The potential is high: state-of-the-art generative models violate kirigami's geometric placement constraints so frequently that their success rates fall below 25\% \cite{felsch2024generative}. 
A geometry-aware surrogate thus holds potential to render the disordered design space navigable.

\begin{figure}[ht!]
\centering
\includegraphics[width=0.75\textwidth]{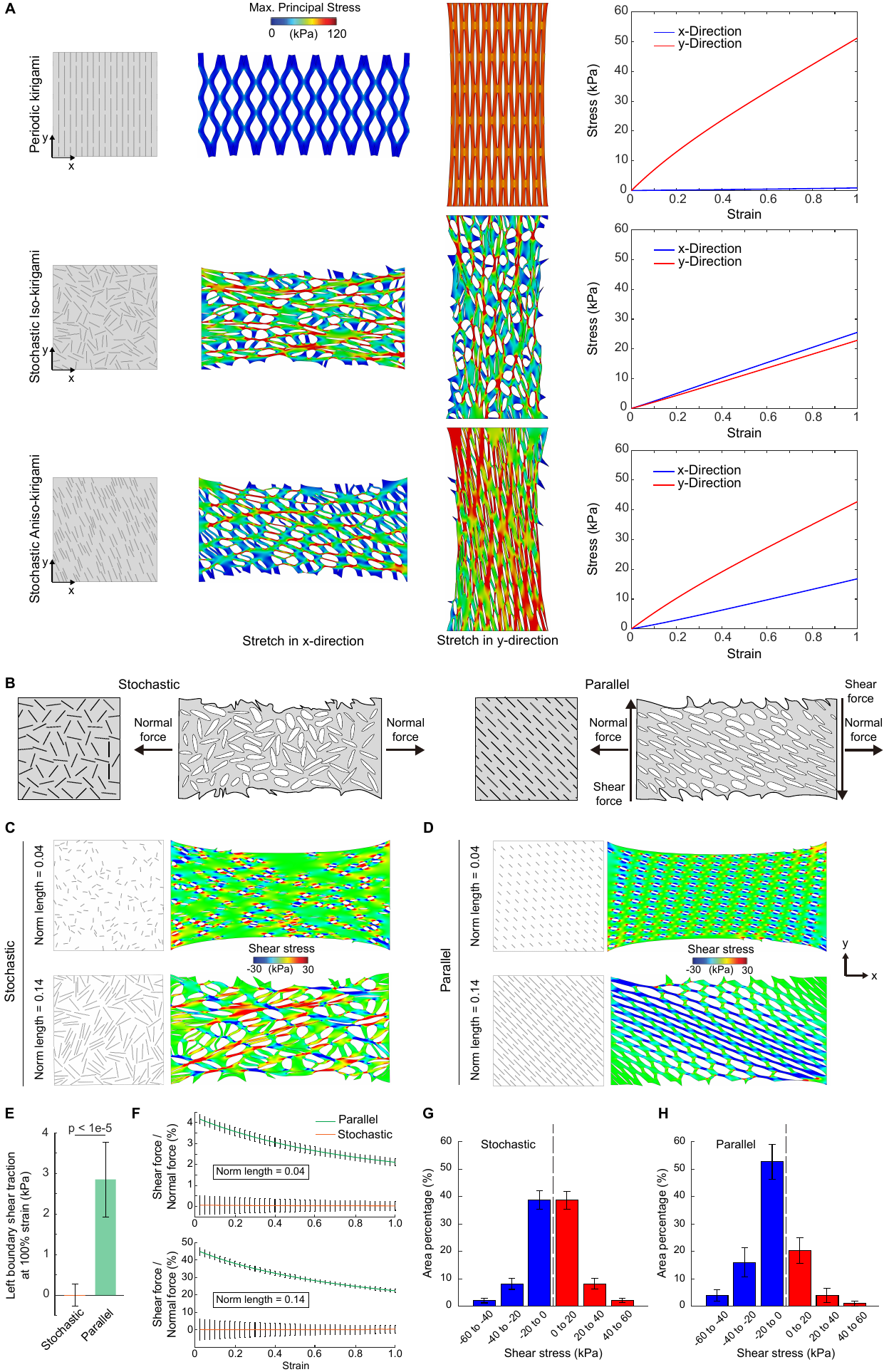}
\caption{\textbf{Stochastic disorder suppresses the extension-shear coupling intrinsic to periodic kirigami, enabling balanced stress states that periodic patterns cannot reach.}
A, Representative periodic and stochastic kirigami architectures. Periodic kirigami exhibits strong directional stiffness, stochastic iso-kirigami produces nearly isotropic behavior, and stochastic aniso-kirigami yields tailored stiffness asymmetry. Finite element simulations of uniaxial stretching in the x and y directions illustrate the corresponding deformation fields, while the accompanying stress–strain curves demonstrate that the mechanical response can be programmed through the cut architecture and orientation distribution. B, Similar stress–strain responses in the x and y directions can be achieved by either stochastic iso-kirigami or periodic kirigami with parallel cuts oriented at 45°. Despite their nearly identical tensile responses, the two architectures exhibit fundamentally different deformation mechanisms, motivating the comparison of their shear behavior.
C, D, Local shear fields for representative stochastic (C) and periodic (D) patterns. Stochastic cuts produce spatially interspersed regions of positive and negative shear that cancel on average, whereas the periodic motif drives a coherent, localized shear band.
E, Shear traction along the loaded boundary at 100\% strain is near zero for stochastic designs but substantial for periodic designs, a direct signature of the absence or presence of net extension-shear coupling.
F, Normalized boundary reaction forces remain negligible for stochastic architectures, consistent with cancellation of opposing shear contributions, whereas periodic architectures retain a persistent directional bias.
G, H, Distributions of local shear stress are symmetric about zero for stochastic patterns (G) but skewed for periodic patterns (H), confirming that the cancellation of shear in disordered designs is a robust statistical property of the field rather than a feature of any single location.}
\label{fig1}
\end{figure}

Finite-element analysis revealed that randomizing cut orientation and placement breaks the correlated rotations that constrain periodic kirigami: the biased shear that periodic motifs concentrate along preferred directions is instead distributed and self-cancelling, leaving near-zero net shear at the boundaries even under 100\% uniaxial strain (Fig.~\ref{fig1}). 
Stochastic kirigami can therefore be designed to extend with almost no parasitic shear, decoupling extension from shear, a pairing essentially inaccessible to periodic patterns. 
Because global behavior no longer follows from a single repeating motif, anisotropy becomes a continuously tunable property set by the statistics of cut orientation and spacing, opening a far broader and smoother region of mechanical response than periodic architectures can reach.

\begin{figure}[ht]
\centering
\includegraphics[width=1.0\textwidth]{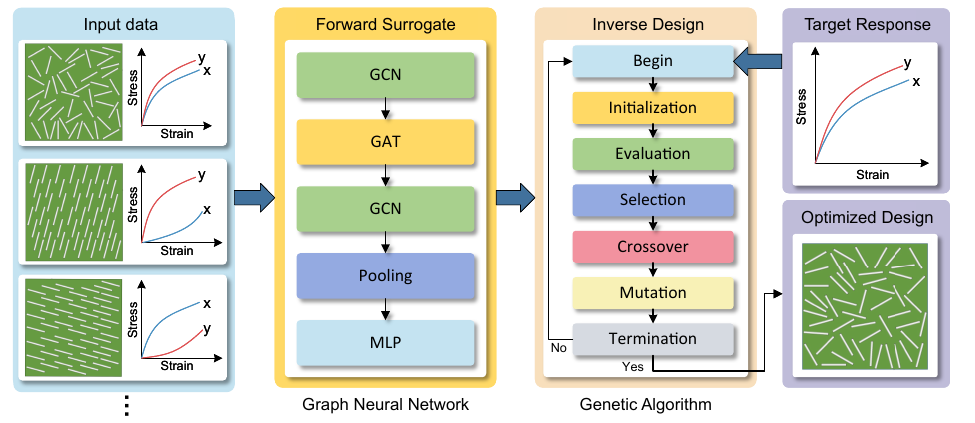}
\caption{\textbf{A graph-neural-network surrogate and genetic-algorithm search form a closed loop for inverse design of stochastic kirigami.} 
Because disordered cut patterns lack a low-dimensional parameterization, the framework instead learns the geometry-to-mechanics map from data and inverts it through optimization. 
Stochastic kirigami with varied cut orientations and spatial distributions produce a broad range of anisotropic stress-strain responses. 
A graph neural network (GNN), trained on finite element data, maps each cut geometry to its full nonlinear response in both loading directions, fast enough to evaluate the many candidate designs generated during optimization. 
A genetic algorithm (GA) exploits this surrogate to search the space of stochastic patterns: it iteratively mutates and recombines candidate designs and scores each prediction against a user-specified target until a matching geometry is identified. 
By replacing direct simulation within the optimization loop, the trained surrogate turns an otherwise intractable inverse problem into a rapid, automated search.}
\label{fig2}
\end{figure}

To turn this principle into a design capability, we developed a framework for forward prediction and inverse design of stochastic kirigami with programmable anisotropy. 
We generated 3,600 finite-element simulations spanning random cut orientations, lengths, and spatial distributions and trained a GNN surrogate to predict the full nonlinear stress-strain response under bidirectional loading (Fig.~\ref{fig2}). 
The surrogate reaches a mean absolute percentage error of roughly 3.5\%, well below a CNN baseline, while training an order of magnitude faster, and we embed it in a genetic algorithm that searches the stochastic cut parameters for patterns matching user-specified mechanical targets. 
Fabricating the optimized designs in silicone elastomer and testing them under uniaxial tension, we find close agreement among target, predicted, and measured responses, validating the complete pipeline from digital design to physical part. 
Together, these results establish controlled disorder as a programmable degree of freedom for architected materials.
When, paired with geometry-aware learning, this yields anisotropic mechanical behavior beyond the reach of periodic design and points toward soft actuators, stretchable electronics, and tissue-interfacing devices such as cardiac patches and skin scaffolds engineered to match the directional mechanics of living tissue.

%
%
\section{Results}\label{sec2}

\subsection{Stochastic disorder decouples extension from shear}\label{subsec2.1}
To explore how cut architecture governs the mechanical behavior of kirigami, we compared three representative designs: a periodic kirigami with parallel cuts, a stochastic design with randomly oriented cuts that produces nearly isotropic behavior (iso-kirigami), and a stochastic design with a preferred cut orientation that generates directional stiffness (aniso-kirigami) (Fig.~\ref{fig1}A). Representative sheets from each class were subjected to uniaxial stretching in both the x and y directions, and the corresponding stress–strain responses were determined by finite element simulations. The results demonstrate that both the type of cut pattern and the distribution of cut orientations provide effective means to tune the nonlinear stress–strain response in each loading direction, enabling responses that range from strongly anisotropic to nearly isotropic. Interestingly, nearly identical stress-strain curves in the x and y directions can be achieved by either stochastic iso-kirigami or periodic kirigami with parallel cuts oriented at 45° (Fig.~\ref{fig1}B). Although these two architectures exhibit similarly isotropic tensile responses, they differ fundamentally in how they deform, revealing that comparable stress–strain behavior does not necessarily imply equivalent mechanics.

The most consequential effect of disorder appears in the shear response. 
In periodic kirigami, the correlated panel rotations generate a coherent, spatially localized shear band, whereas stochastic patterns produce regions of positive and negative shear that are interspersed throughout the sheet and cancel on average (Fig.~\ref{fig1}C,D). 
This local cancellation has a direct macroscopic signature: at 100\% strain, the shear traction along the loaded boundary is near zero for stochastic designs but substantial for periodic ones (Fig.~\ref{fig1}E), and the corresponding normalized boundary reaction forces remain negligible for stochastic architectures while periodic designs retain a persistent directional bias (Fig.~\ref{fig1}F). 
Across the field, the distribution of local shear stress is symmetric about zero for stochastic patterns but skewed for periodic ones (Fig.~\ref{fig1}G,H). 
Together, these results show that a stochastic sheet can be stretched along one axis while developing essentially no net shear, a decoupling of extension from shear that the kinematic constraints of a periodic motif preclude.

This behavior has two implications for design. 
First, it grants access to mechanical responses that are difficult to obtain from periodic systems, such as the combination of large extensibility with near-zero net shear. 
Second, because the global response no longer derives from a single repeating unit, anisotropy becomes a quantity that can be tuned continuously through the statistics of cut orientation and spacing rather than selected from a discrete set of motifs. 
In effect, disorder averages local stress states instead of amplifying them, yielding a more balanced response at the scale of the whole sheet. What disorder does not provide is a simple description of how a given cut pattern maps to this behavior: unlike periodic lattices, stochastic architectures have no low-dimensional parameterization through which a target response can be located. 
We therefore turned to a data-driven surrogate to make this expanded design space navigable.

\begin{figure}[ht]
\centering
\includegraphics[width=1.0\textwidth]{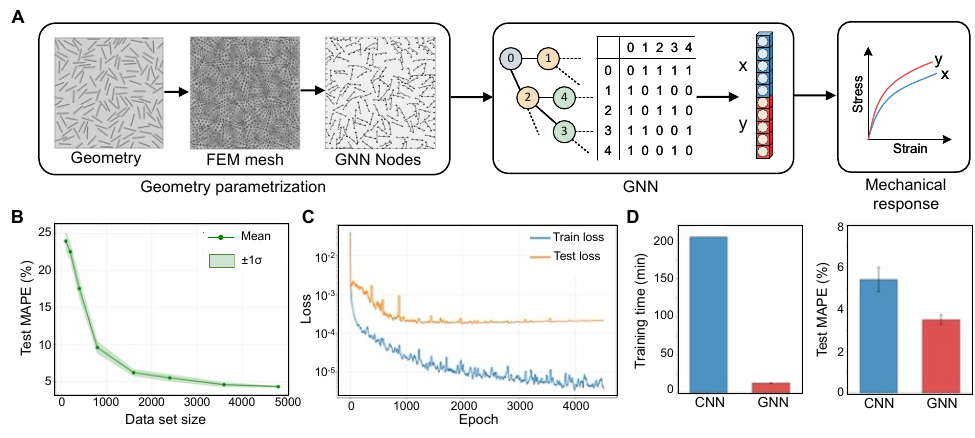}
\caption{\textbf{A graph neural network surrogate predicts the nonlinear anisotropic response of stochastic kirigami.}
A, Model pipeline: each kirigami pattern is converted into a node-edge graph by extracting the geometric points from the ABAQUS input file. The extracted points form the graph nodes, while undirected edges connect consecutive points along each cut, preserving the cut geometry and topology. The graph neural network (GNN) maps this representation to the full stress-strain response in both the $x$ and $y$ loading directions.
B, Test-set mean absolute percentage error (MAPE; mean $\pm 1\sigma$) falls as the training set grows and plateaus beyond roughly 3,500 geometries.
C, Training and test loss curves decay smoothly and together, indicating stable convergence and negligible overfitting.
D, Head-to-head comparison of GNN and CNN surrogates trained on the same data: the GNN is both more accurate (MAPE $3.50 \pm 0.24\%$ versus $5.42 \pm 0.58\%$, $p < 0.01$) and roughly fifteen times faster to train ($13.3 \pm 0.5$ versus $204.7 \pm 0.1$ min, $p < 0.001$).}
\label{fig3}
\end{figure}

\subsection{A graph neural network learns the structure-property map of disordered kirigami}\label{subsec2.2}
A surrogate able to stand in for finite element simulation inside an optimization loop would render this space searchable, if it could learn the geometry-to-mechanics map both accurately and cheaply. We built such a surrogate as a graph neural network (GNN), choosing its representation to match the physics of the problem (Fig.~\ref{fig3}A). 
Each design was converted into a node-edge graph that encodes the cuts and their geometric connectivity, so that the cut interactions and deformation-relevant structure governing the response are represented directly (See \textbf{Materials and Methods}). 
This is the information that is blurred or discarded when a pattern is rasterized into a pixel image and processed by a convolutional neural network (CNN). From this graph alone, the trained GNN predicts the full nonlinear stress-strain curve in both the $x$ and $y$ directions.

Test accuracy improved as the training set grew from 250 to 5,000 geometries and plateaued beyond roughly 3,500 (Fig.~\ref{fig3}B); we therefore trained on 3,600 finite element simulations, which met our target of below 5\% error. 
Training was stable, with training and test losses decaying smoothly and in step, indicating strong generalization and negligible overfitting across the diverse stochastic geometries (Fig.~\ref{fig3}C). 
A direct comparison confirmed that the graph representation is the better choice on both of the axes that matter for an in-loop surrogate (Fig.~\ref{fig3}D): trained on the same data, the GNN reached a lower error than the CNN (mean absolute percentage error, MAPE, of $3.50 \pm 0.24\%$ versus $5.42 \pm 0.58\%$, $p < 0.01$) while training roughly fifteen times faster ($13.3 \pm 0.5$ versus $204.7 \pm 0.1$ min, $p < 0.001$), avoiding the heavy convolutional processing that high-resolution pattern images demand.

\begin{figure}[ht]
\centering
\includegraphics[width=1.0\textwidth]{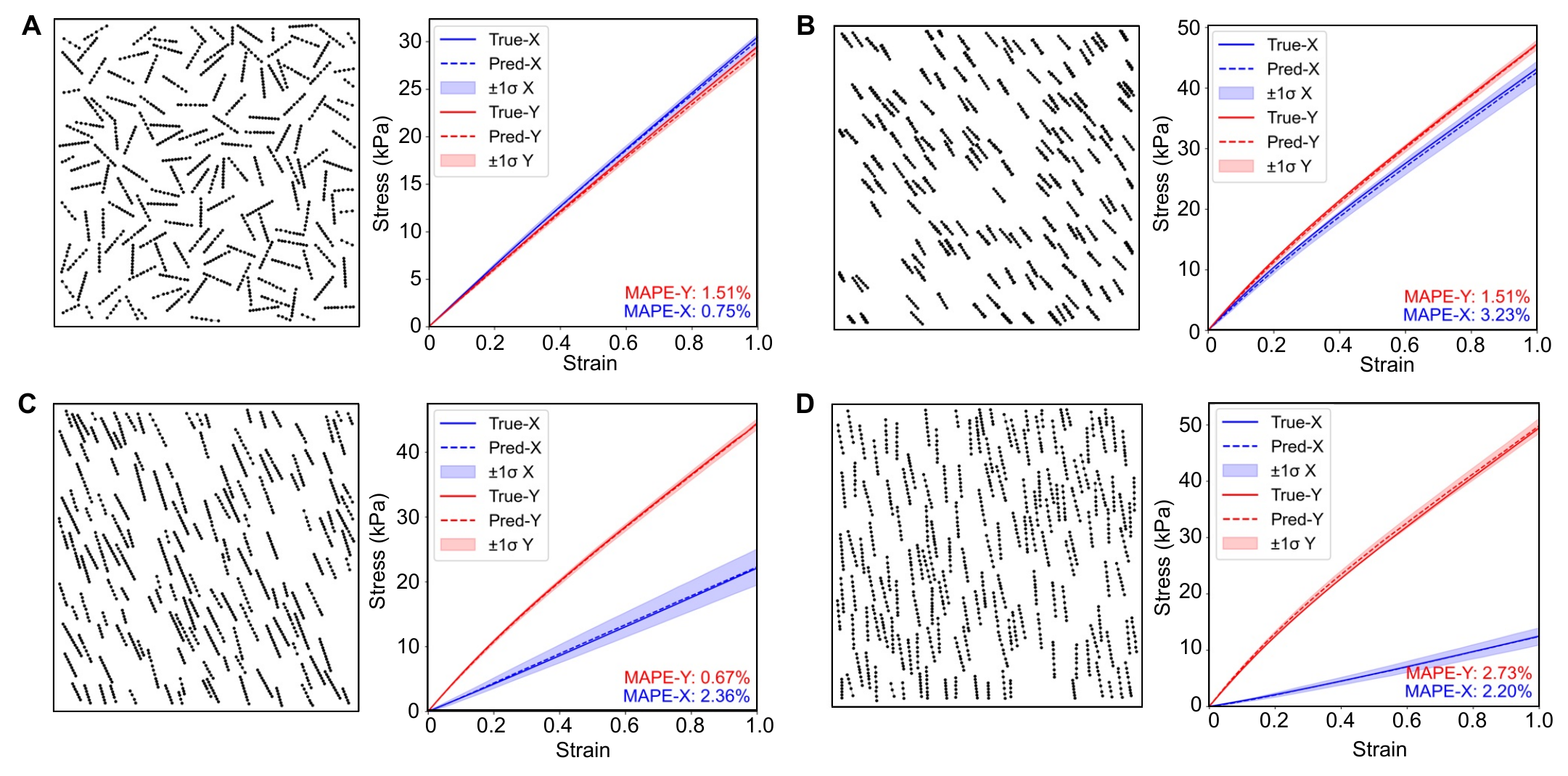}
\caption{\textbf{GNN predictions match the finite element ground truth across diverse kirigami designs.} 
A-D, For four representative designs with differing cut geometries, the kirigami pattern is shown beside its stress-strain response in both loading directions, comparing finite element ground truth (solid) with GNN predictions (dashed). The surrogate reproduces the nonlinear response in every case, with MAPE below 3.5\%. 
Shaded bands denote $\pm 1\sigma$ variability across seven independently trained GNN surrogates.}\label{fig4}
\end{figure}

The surrogate reproduced the response of held-out designs with high fidelity, tracking the ground-truth finite element curves in both loading directions across patterns of widely varying geometry, with MAPE below 3.5\% in every case (Fig.~\ref{fig4}). 
Narrow variability across seven independently trained GNNs indicates that this accuracy is robust to initialization rather than the product of a single fortuitous fit. 
With a fast, accurate, and reliable map from geometry to mechanical response in hand, we turned to inverting the design problem.

\subsection{Surrogate-guided optimization discovers patterns matching prescribed targets}\label{subsec2.3}
\begin{figure}[htbp]
\centering
\includegraphics[width=0.84\textwidth]{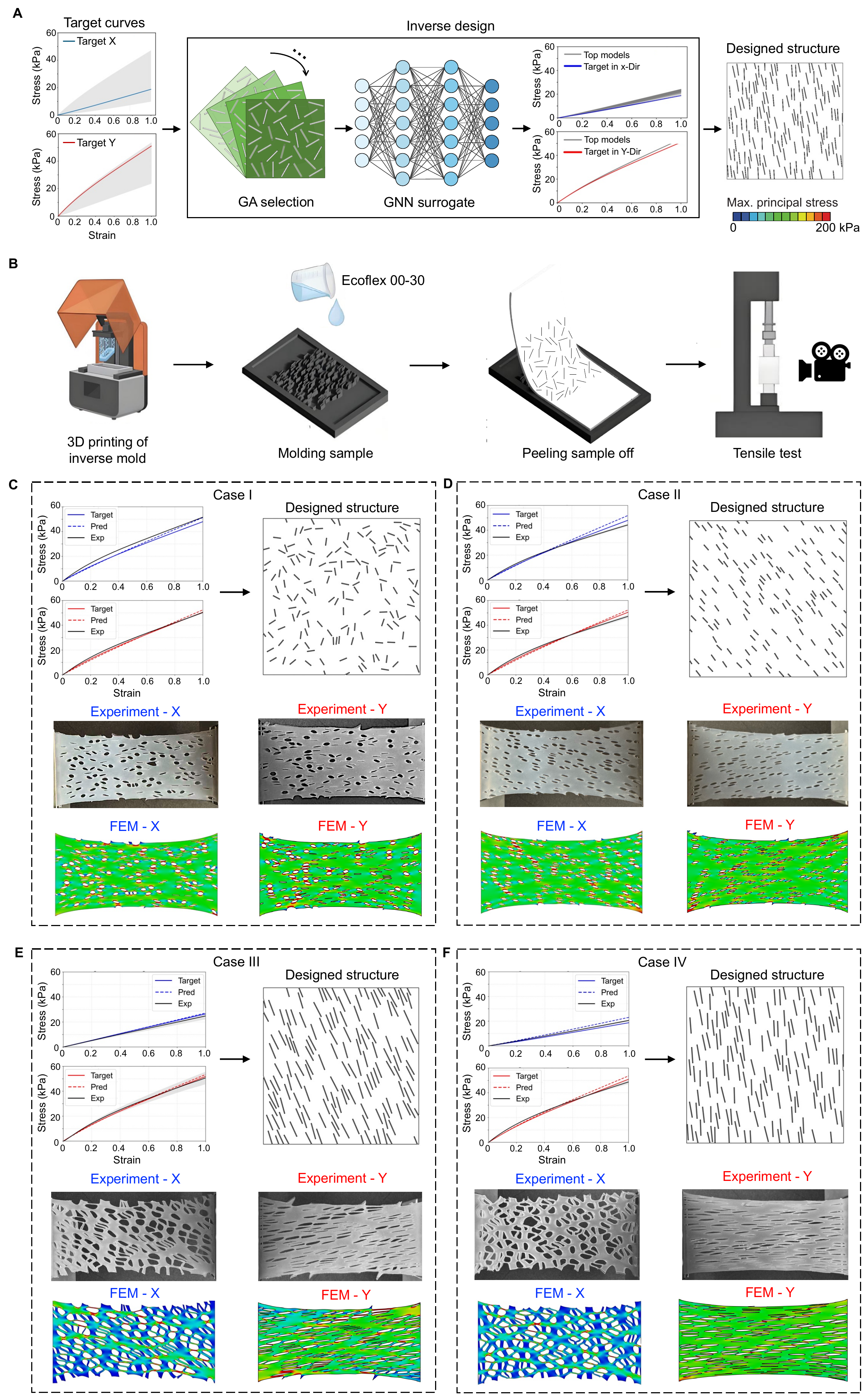}
\caption{\textbf{Inverse design and experimental validation of stochastic kirigami with prescribed anisotropic responses.}
A, Inverse-design workflow. 
A target response, constructed by blending three curves from the training set so that it lies within the achievable range, defines the objective. 
A genetic algorithm (GA) evolves candidate patterns while the GNN scores each predicted response against the target, converging on a geometry whose predicted curves match the target in both loading directions.
B, Fabrication and testing.
Inverse molds of the optimized designs were 3D-printed, filled with Ecoflex 00-30, cured at room temperature, demolded, and loaded in uniaxial tension.
C-F, For four representative inverse-designed samples, the target (colored solid), GNN-predicted (colored dashed), and measured (black solid) stress-strain curves agree in both loading directions. Insets compare a photograph of each sample at 100\% strain with the finite element field of maximum principal stress.}
\label{fig5}
\end{figure}

Inverse design reverses the forward map: instead of predicting the response of a given pattern, we prescribe a target response and search for a disordered geometry that produces it. 
We embedded the trained GNN in a genetic algorithm (GA) that takes target stress-strain curves in both the $x$ and $y$ directions and evolves a population of candidate designs toward them (Fig.~\ref{fig5}A). 
To ensure each target lay within the achievable response space, we constructed it by blending three curves drawn from the training set. 
The GA operated not on individual cut placements but on the parametric description of the disorder: the orientation ranges, spacing, and density that set the statistics of the cut field. 
Thus, at each generation it mutates and recombines these parameters, generates the corresponding pattern, and scores the GNN-predicted response against the target, continuing until the best candidate's predicted MAPE falls below 5\% or a maximum of 20 generations is reached (Materials and Methods).

Because the surrogate returns each evaluation in a fraction of the time a finite element solve would take, the GA can explore a large design space and converge on geometries whose predicted responses closely track the prescribed targets in both loading directions (Fig.~\ref{fig5}A).
Searching over the statistics of the cut field rather than placing individual cuts carries a further benefit: every candidate the GA proposes is, by construction, a physically valid pattern, so the search never wastes effort on infeasible geometries. 
To confirm that these predicted matches survive the passage from digital design to physical part, we fabricated and tested the optimized structures.

\subsection{Fabricated samples confirm the predicted responses}\label{subsec2.4}
Inverse molds of the four optimized designs were 3D-printed, filled with a soft silicone elastomer (Ecoflex 00-30, Smooth-On), cured at room temperature, and demolded; the resulting specimens were loaded in uniaxial tension to measure their stress-strain response (Fig.~\ref{fig5}B).
For all four samples, the measured curves agreed with both the surrogate prediction and the target in each loading direction, reproducing the same nonlinear trends and direction-dependent anisotropy (Fig.~\ref{fig5}C-F).

The agreement extended beyond the bulk response to the deformation itself: at 100\% strain, the deformed shapes observed experimentally matched those predicted by the finite element model, shown alongside the corresponding fields of maximum principal stress (insets, Fig.~\ref{fig5}C-F). 
The minor deviations that remained are consistent with fabrication tolerances and batch-to-batch variability in the elastomer, and do not alter the agreement in trend or anisotropy. 
Taken together with results of the preceding sections, these results close the loop from a prescribed target to a physically realized part, confirming that a graph-neural-network surrogate coupled to evolutionary search can design stochastic kirigami with specified anisotropic responses.
%
%
\section{Discussion}\label{sec3}

\subsection{Disorder-enabled design and geometry-aware learning}
Results show that disorder need not be treated as a defect to be minimized, but instead can serve as a design variable to be exploited \cite{zaiser2023disordered}. 
Periodic architectures sample a restricted and discrete subspace of mechanical responses \cite{yu2025expansion}, whereas stochastic kirigami access a broader and more continuous space, including the extension-shear decoupling that periodic motifs cannot achieve. 
This expanded tunability echoes settings in which randomness enhances fracture toughness \cite{fulco2025disorder} and relaxes strength-toughness trade-offs \cite{choukir2025disorder}.
It also parallels  strategies by which biological tissues build compliance \cite{genin2009functional, cs2025molecular}, toughness \cite{hu2015stochastic, genin2017unification}, and directional stiffness \cite{jansen2018role, annaidh2012characterization} from disordered networks. 
Our framework offers a route to harness the same principles in engineered systems.

Exploiting disorder this way was possible because the surrogate's representation matches the physics of the problem. 
Finite element results show that kirigami mechanics are governed by how neighboring cuts interact, how panels rotate under constraint, and how deformation propagates through the network, all features that depend on connectivity rather than absolute spatial position. 
By encoding cuts as nodes and their interactions as edges, the GNN captures these governing mechanisms directly, whereas image-based representations must infer them from pixelated geometry. 
This alignment between representation and behavior accounts for both the higher accuracy and the lower training cost, particularly for stochastic systems in which connectivity varies from one realization to the next, and it is consistent with reports that graph methods outperform grid-based ones for polycrystalline materials \cite{dai2021graph} and disordered lattices \cite{zheng2023unifying}. 
The roughly fifteenfold reduction in training time (Fig.~\ref{fig3}D) makes such surrogates especially well suited to the iterative optimization that searching a large stochastic design space demands.

\subsection{Practical implications and future directions}
The agreement among target, predicted, and measured stress-strain curves confirms the practical value of the approach. 
The surrogate captures the geometric nonlinearities essential to kirigami mechanics, and the genetic algorithm searches parametric stochastic patterns rather than optimizing individual cut placements, sidestepping the constraint violations that hamper generative approaches \cite{felsch2024generative}. 
The framework nonetheless has limits. It assumes planar, quasi-static loading and does not capture out-of-plane buckling or inelastic behavior, and extrapolation beyond the training distribution remains unreliable, a shortcoming that physics-informed architectures may help address \cite{bastek2023inverse}.

Several extensions follow naturally. 
Uncertainty quantification would identify designs that are robust to fabrication variability, which matters for biomedical applications with patient-specific anatomy. 
Multi-objective optimization could balance anisotropic stiffness against porosity and stress concentration. 
Differentiable surrogates \cite{dold2023differentiable} or reinforcement-learning policies \cite{maurizi2025designing} could speed convergence on demanding inverse problems. 
Beyond kirigami, the methodology generalizes to any architected material that lacks an analytical parameterization, including origami structures, lattice scaffolds, and network materials, all of which share the property that response emerges from discrete elements and their connectivity. 
With suitable training data, the same pipeline could design cardiac patches matched to myocardial orthotropy \cite{olvera2020electroconductive}, skin scaffolds reproducing the anisotropy of Langer lines \cite{annaidh2012characterization}, or conformable sensors that retain function under multiaxial strain \cite{rogers2010materials, xue2020mechanically}.
 %
 %
\section{Conclusions}\label{sec4}

We have shown that controlled disorder is a design variable for kirigami metamaterials, granting access to anisotropic responses beyond the reach of periodic architectures, including the decoupling of extension from shear. 
Because disordered cut patterns lack a simple parameterization, we made this space navigable with a geometry-aware graph neural network that maps cut topology to the full nonlinear, bidirectional stress-strain response far more efficiently than image-based models, coupled to a genetic algorithm for inverse design. 
Fabricated elastomer samples matched their prescribed targets in both loading directions, closing the loop from design to physical part. 
The approach generalizes to any architected material whose response emerges from discrete elements and their connectivity, establishing disorder, paired with geometry-aware learning, as a programmable route to anisotropic mechanical function.

\backmatter

\section*{Materials and Methods}\label{sec5}

\subsection*{Training data generation from finite element simulations}\label{subsec5.1}

Stochastic kirigami microstructures used for training data were generated automatically through a custom Python script that constructs non-overlapping cut patterns under predefined geometric constraints. The base domain was a 50 mm × 50 mm rectangular shell. First, a total of 1,200 candidate center points were generated by iterative rejection sampling to ensure a minimum point-point separation of 0.5 mm. Each cut was modeled as a thin rectangular slit with a fixed width of 0.2 mm. The slit length for each sample was drawn randomly between 2-7 mm. Slit orientations were extracted from predefined angular ranges (e.g., 0-15°, 15-30°, 30-45°, or 0-360°) to control for directional deviations across different datasets. For each placement attempt, the global coordinates of the slit were computed via rigid-body transformation and evaluated using two geometric filters: (1) an exact segment-segment intersection test against edges of all previously placed cuts, and (2) verification that the minimum distance between the new slit and all existing slits exceeded 0.5 mm, computed via bidirectional point-to-segment distance calculations. Cuts passing both criteria were accepted until reaching 150 cuts per pattern. Once all cuts were accepted, the pattern was applied to the base sheet through Boolean subtraction. 
The final kirigami part was then meshed using S4R shell elements with a global mesh size of 0.25 mm. A mesh convergence study was performed to confirm the adequacy of this resolution (Fig.~\ref{fig:SI1}). 

The material was modeled using a three-term Ogden hyperelastic model calibrated from uniaxial tests of uncut Ecoflex 00-30 dogbone samples, with strain energy density given by:

\begin{equation}
W = \sum_{i=1}^{3} \frac{\mu_i}{\alpha_i} 
\left( \lambda_1^{\alpha_i} + \lambda_2^{\alpha_i} + \lambda_3^{\alpha_i} - 3 \right).
\end{equation}

The material parameters are given by 
$\mu_1 = 1164.77~\text{kPa}$, $\alpha_1 = 0.03$, 
$\mu_2 = 5.31~\text{kPa}$, $\alpha_2 = 3.48$, 
and $\mu_3 = 267.86~\text{kPa}$, $\alpha_3 = 0.03$. 
The corresponding material fit is shown in Fig.~\ref{fig:SI2}. 

For mechanical loading of the cut sheets, one boundary of the sheet was fully constrained using fixed boundary conditions, while the opposite boundary was subjected to a prescribed horizontal displacement corresponding to 100\% engineering strain. Reaction forces and displacements were recorded at the loaded boundary throughout the simulation, generating the force-displacement curve in the horizontal direction (x-direction). This same uniaxial stretching protocol was also performed in the perpendicular direction to obtain the force-displacement curve in the vertical direction (y-direction). Each kirigami microstructure thus produced paired force--displacement curves for both x and y loading directions, forming the complete training dataset for the GNN surrogate model.

\subsection*{Graph neural network (GNN) surrogate for forward prediction}

The forward surrogate model was implemented as a GNN designed to predict the stress-strain response of a kirigami structure directly from its cut-network graph. Each training sample comprised (i) a graph representation of the microstructure and (ii) stress-strain curves in both the $x$- and $y$-loading directions. The graph representation was extracted directly from the Abaqus input files by identifying the geometric points defining each kirigami cut. Each extracted point was treated as a graph node, with its in-plane coordinates $(x,y)$ used as the node features. An undirected edge was assigned between consecutive points belonging to the same cut, such that each cut was represented as a connected polyline and the complete kirigami pattern as the collection of all cut subgraphs. For the datasets considered in this study, each graph contained approximately 1,200 nodes and 1,050 edges. This graph representation preserves the spatial arrangement and topology of the kirigami architecture, enabling the GNN to learn the relationship between cut geometry and mechanical response.
For each sample, both directional responses were linearly interpolated to 30 uniformly spaced strain levels, concatenated into a 60-dimensional stress vector, and normalized by its maximum absolute value. To enforce rotational consistency, every sample was paired with an augmented version generated by rotating all node coordinates by 90° around the origin while preserving edge connectivity, and by swapping the corresponding stress vector. Following augmentation, the dataset was split into approximately 80\% training and 20\% testing.

The GNN architecture follows a GCN-GAT-GCN message-passing sequence. The first graph convolutional layer (GCNConv) maps the 2D node coordinates into a 128-dimensional latent space and aggregates information from local neighborhoods. The second layer was a four-head graph attention module (GATConv) with non-concatenated outputs, producing a 128-dimensional embedding. A third GCN layer further propagates these attention-refined features across the graph, retaining a 128-dimensional node of representation. Following the message-passing stages, a global mean-pooling operation compresses all node embeddings into a single 128-dimensional graph-level descriptor that summarizes the topology, spatial arrangement, and density of cuts. This descriptor was then processed by a multilayer perceptron consisting of a fully connected layer reducing the dimensionality from 128 to 84 with a ReLU activation, followed by a linear output layer that maps to the 60 components of the resampled stress vector.

Model training proceeds for up to 2,000 epochs using the Adam optimizer with a learning rate of $2 \times 10^{-3}$ and a mean-squared-error (MSE) loss between predicted and target vectors, using mini-batches of 64 graphs. After convergence, the trained weights were saved to a standalone model file, yielding a fast and accurate surrogate capable of mapping any input kirigami graph to its corresponding mechanical responses.

\subsection*{Genetic algorithm (GA) for inverse design}

The inverse-design problem was formulated as a genetic algorithm (GA) that searches over kirigami microstructure parameters, with the trained GNN surrogate replacing computationally expensive FEM simulations during evaluation. Each individual encodes a set of geometric and stochastic ``genes,'' including the random seed governing cut placement, the minimum cut separation, cut thickness, bounds on cut height, and the angular range that constrains allowable cut orientations. To promote systematic exploration of orientation space, the population was partitioned into fixed orientation ``buckets'' (e.g., $0^\circ$--$15^\circ$, $15^\circ$--$30^\circ$, $30^\circ$--$45^\circ$, and $0^\circ$--$360^\circ$), with each population member assigned to one bucket throughout evolution. This strategy prevents early convergence to a narrow subset of orientations and ensures that a broad spectrum of anisotropic patterns is explored during optimization.

For each individual, an Abaqus/CAE script procedurally generates a non-overlapping set of rectangular cuts within the base sheet using the prescribed gene parameters. The script samples a constant cut height within the specified bounds, enforces geometric constraints to prevent overlap, applies uniaxial loading boundary conditions, and exports the mesh geometry as an Abaqus input file. The resulting boundary coordinates are then parsed and clustered into a reduced cut-network graph, which serves as input to the trained GNN surrogate.

During optimization, the surrogate predicts the 60-component stress vector for each candidate design, and the predictions are compared with the target curves in both loading directions. The fitness is defined as the weighted sum of direction-wise mean squared error (MSE) values, with an additional aggregate MSE criterion used to filter out poorly performing individuals. After evaluation, the best-performing designs are archived, and the next generation is formed using tournament selection, crossover, and constrained mutation applied within allowable gene ranges and orientation buckets. When surrogate-predicted losses exceed a threshold, new random seeds are introduced to enhance diversity and avoid entrapment in local minima. This surrogate-assisted GA loop is iterated until convergence criteria are satisfied, defined by the best individual achieving a mean absolute percentage error (MAPE) below $5\%$.

\subsection*{Experimental verification}\label{subsec5.4}

Three-dimensional geometries with a uniform thickness of 2 mm were imported into SolidWorks to generate corresponding inverse molds designed to accurately capture all cut shapes and geometric features. The molds were fabricated using a high-resolution 3D printer (FormLabs 3, FormLabs, Massachusetts, US) to preserve fine pattern details. Ecoflex-0030, a platinum-cured silicone elastomer with high stretchability and compliance, was prepared by mixing its base (Part A) and curing agent (Part B) at the 1:1 ratio, followed by vacuum degassing to remove trapped air bubbles. The degassed mixture was slowly poured into the printed molds to ensure complete filling of the patterned regions. The molds were left to cure at room temperature overnight to allow full crosslinking of the elastomer. After curing, the cast samples were carefully demolded to obtain the final patterned kirigami specimens. These specimens were then mounted in a uniaxial testing machine (Model 5583, Instron, Massachusetts, US) and subjected to tensile loading to characterize their mechanical responses for comparison with surrogate predictions and target curves.

\begin{appendices}

\section{Supplementary figures}\label{secA1}

\begin{figure}[htbp]
\centering
\includegraphics[width=0.5\linewidth]{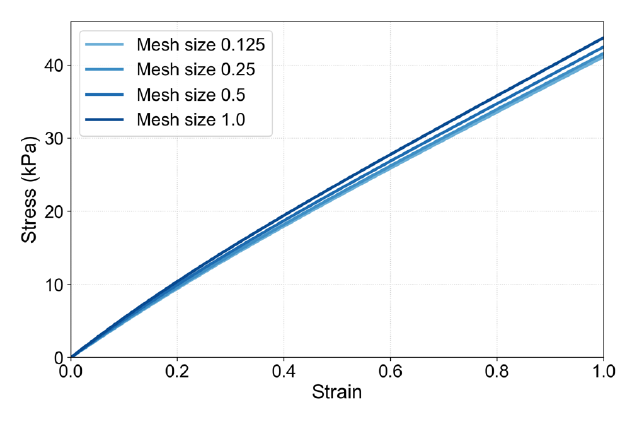}
\caption{Mesh convergence analysis for the finite element simulation.}
\label{fig:SI1}
\end{figure}

\begin{figure}[htbp]
\centering
\includegraphics[width=0.5\linewidth]{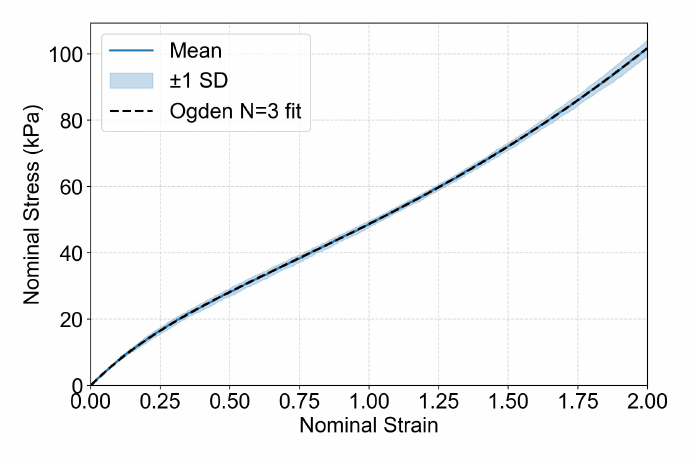}
\caption{Three-term Ogden hyperelastic model calibration using experimental stress–strain data from uncut Ecoflex-0030 dogbone specimens.}
\label{fig:SI2}
\end{figure}

\end{appendices}

\subsection*{Acknowledgements}
This work was funded by the NSF through grants OIA-2219142 and CMMI 1548571, and by the NIH through grants R01 AR084243, R01 DK131177, R01 AR077793, R01 HL159094, and R01 AR084243. This work was also partially supported by an NSF Accelerating Research Translation cooperative agreement (TIP-2331429) and the NJIT Center for Translational Research. The opinions, findings, and conclusions, or recommendations expressed are those of the authors and do not necessarily reflect the views of the NSF and NIH.

\subsection*{Declarations}
The authors declare no competing interests.

\subsection*{Data Availability}
All codes and simulation dataset used in this study are available at \url{https://github.com/jinhanxun/Kirigami_GNN}.

\bibliography{sn-bibliography}

\begin{thebibliography}{10}
\expandafter\ifx\csname url\endcsname\relax
  \def\url#1{\burl{#1}}\fi
\expandafter\ifx\csname urlprefix\endcsname\relax\def\urlprefix{URL }\fi
\providecommand{\bibinfo}[2]{#2}
\providecommand{\eprint}[2][]{\url{#2}}
\providecommand{\doi}[1]{\url{https://doi.org/#1}}
\bibcommenthead

\bibitem{bertoldi2017flexible}
\bibinfo{author}{Bertoldi, K.}, \bibinfo{author}{Vitelli, V.}, \bibinfo{author}{Christensen, J.} \& \bibinfo{author}{Van~Hecke, M.}
\newblock \bibinfo{title}{Flexible mechanical metamaterials}.
\newblock \emph{\bibinfo{journal}{Nature Reviews Materials}} \textbf{\bibinfo{volume}{2}}, \bibinfo{pages}{1--11} (\bibinfo{year}{2017}).

\bibitem{jin2024mechanical}
\bibinfo{author}{Jin, H.} \& \bibinfo{author}{Espinosa, H.~D.}
\newblock \bibinfo{title}{Mechanical metamaterials fabricated from self-assembly: A perspective}.
\newblock \emph{\bibinfo{journal}{Journal of Applied Mechanics}} \textbf{\bibinfo{volume}{91}}, \bibinfo{pages}{040801} (\bibinfo{year}{2024}).

\bibitem{jiao2023mechanical}
\bibinfo{author}{Jiao, P.}, \bibinfo{author}{Mueller, J.}, \bibinfo{author}{Raney, J.~R.}, \bibinfo{author}{Zheng, X.} \& \bibinfo{author}{Alavi, A.~H.}
\newblock \bibinfo{title}{Mechanical metamaterials and beyond}.
\newblock \emph{\bibinfo{journal}{Nature communications}} \textbf{\bibinfo{volume}{14}}, \bibinfo{pages}{6004} (\bibinfo{year}{2023}).

\bibitem{surjadi2019mechanical}
\bibinfo{author}{Surjadi, J.~U.} \emph{et~al.}
\newblock \bibinfo{title}{Mechanical metamaterials and their engineering applications}.
\newblock \emph{\bibinfo{journal}{Advanced Engineering Materials}} \textbf{\bibinfo{volume}{21}}, \bibinfo{pages}{1800864} (\bibinfo{year}{2019}).

\bibitem{jin2026situ}
\bibinfo{author}{Jin, H.} \emph{et~al.}
\newblock \bibinfo{title}{In situ mechanical characterization of functional and architected materials}.
\newblock \emph{\bibinfo{journal}{Nature Materials}} \bibinfo{pages}{1--19} (\bibinfo{year}{2026}).

\bibitem{jin2025characterization}
\bibinfo{author}{Jin, H.} \emph{et~al.}
\newblock \bibinfo{title}{Characterization and inverse design of stochastic mechanical metamaterials using neural operators}.
\newblock \emph{\bibinfo{journal}{Advanced Materials}} \textbf{\bibinfo{volume}{37}}, \bibinfo{pages}{2420063} (\bibinfo{year}{2025}).

\bibitem{shyu2015kirigami}
\bibinfo{author}{Shyu, T.~C.} \emph{et~al.}
\newblock \bibinfo{title}{A kirigami approach to engineering elasticity in nanocomposites through patterned defects}.
\newblock \emph{\bibinfo{journal}{Nature materials}} \textbf{\bibinfo{volume}{14}}, \bibinfo{pages}{785--789} (\bibinfo{year}{2015}).

\bibitem{blees2015graphene}
\bibinfo{author}{Blees, M.~K.} \emph{et~al.}
\newblock \bibinfo{title}{Graphene kirigami}.
\newblock \emph{\bibinfo{journal}{Nature}} \textbf{\bibinfo{volume}{524}}, \bibinfo{pages}{204--207} (\bibinfo{year}{2015}).

\bibitem{zhang2015mechanically}
\bibinfo{author}{Zhang, Y.} \emph{et~al.}
\newblock \bibinfo{title}{A mechanically driven form of kirigami as a route to 3d mesostructures in micro/nanomembranes}.
\newblock \emph{\bibinfo{journal}{Proceedings of the National Academy of Sciences}} \textbf{\bibinfo{volume}{112}}, \bibinfo{pages}{11757--11764} (\bibinfo{year}{2015}).

\bibitem{meng2022kirigami}
\bibinfo{author}{Meng, K.} \emph{et~al.}
\newblock \bibinfo{title}{Kirigami-inspired pressure sensors for wearable dynamic cardiovascular monitoring}.
\newblock \emph{\bibinfo{journal}{Advanced Materials}} \textbf{\bibinfo{volume}{34}}, \bibinfo{pages}{2202478} (\bibinfo{year}{2022}).

\bibitem{brooks2022kirigami}
\bibinfo{author}{Brooks, A.~K.}, \bibinfo{author}{Chakravarty, S.}, \bibinfo{author}{Ali, M.} \& \bibinfo{author}{Yadavalli, V.~K.}
\newblock \bibinfo{title}{Kirigami-inspired biodesign for applications in healthcare}.
\newblock \emph{\bibinfo{journal}{Advanced Materials}} \textbf{\bibinfo{volume}{34}}, \bibinfo{pages}{2109550} (\bibinfo{year}{2022}).

\bibitem{rafsanjani2017buckling}
\bibinfo{author}{Rafsanjani, A.} \& \bibinfo{author}{Bertoldi, K.}
\newblock \bibinfo{title}{Buckling-induced kirigami}.
\newblock \emph{\bibinfo{journal}{Physical review letters}} \textbf{\bibinfo{volume}{118}}, \bibinfo{pages}{084301} (\bibinfo{year}{2017}).

\bibitem{genin2017unification}
\bibinfo{author}{Genin, G.~M.} \& \bibinfo{author}{Thomopoulos, S.}
\newblock \bibinfo{title}{Unification through disarray}.
\newblock \emph{\bibinfo{journal}{Nature materials}} \textbf{\bibinfo{volume}{16}}, \bibinfo{pages}{607--608} (\bibinfo{year}{2017}).

\bibitem{golman2021toughening}
\bibinfo{author}{Golman, M.} \emph{et~al.}
\newblock \bibinfo{title}{Toughening mechanisms for the attachment of architectured materials: The mechanics of the tendon enthesis}.
\newblock \emph{\bibinfo{journal}{Science Advances}} \textbf{\bibinfo{volume}{7}}, \bibinfo{pages}{eabi5584} (\bibinfo{year}{2021}).

\bibitem{zaiser2023disordered}
\bibinfo{author}{Zaiser, M.} \& \bibinfo{author}{Zapperi, S.}
\newblock \bibinfo{title}{Disordered mechanical metamaterials}.
\newblock \emph{\bibinfo{journal}{Nature Reviews Physics}} \textbf{\bibinfo{volume}{5}}, \bibinfo{pages}{679--688} (\bibinfo{year}{2023}).

\bibitem{annaidh2012characterization}
\bibinfo{author}{Annaidh, A.~N.}, \bibinfo{author}{Bruy{\`e}re, K.}, \bibinfo{author}{Destrade, M.}, \bibinfo{author}{Gilchrist, M.~D.} \& \bibinfo{author}{Ott{\'e}nio, M.}
\newblock \bibinfo{title}{Characterization of the anisotropic mechanical properties of excised human skin}.
\newblock \emph{\bibinfo{journal}{Journal of the mechanical behavior of biomedical materials}} \textbf{\bibinfo{volume}{5}}, \bibinfo{pages}{139--148} (\bibinfo{year}{2012}).

\bibitem{tueni2023structural}
\bibinfo{author}{Tueni, N.}, \bibinfo{author}{Allain, J.-M.} \& \bibinfo{author}{Genet, M.}
\newblock \bibinfo{title}{On the structural origin of the anisotropy in the myocardium: Multiscale modeling and analysis}.
\newblock \emph{\bibinfo{journal}{Journal of the mechanical behavior of biomedical materials}} \textbf{\bibinfo{volume}{138}}, \bibinfo{pages}{105600} (\bibinfo{year}{2023}).

\bibitem{jansen2018role}
\bibinfo{author}{Jansen, K.~A.} \emph{et~al.}
\newblock \bibinfo{title}{The role of network architecture in collagen mechanics}.
\newblock \emph{\bibinfo{journal}{Biophysical journal}} \textbf{\bibinfo{volume}{114}}, \bibinfo{pages}{2665--2678} (\bibinfo{year}{2018}).

\bibitem{chen2020deterministic}
\bibinfo{author}{Chen, S.}, \bibinfo{author}{Choi, G.~P.} \& \bibinfo{author}{Mahadevan, L.}
\newblock \bibinfo{title}{Deterministic and stochastic control of kirigami topology}.
\newblock \emph{\bibinfo{journal}{Proceedings of the National Academy of Sciences}} \textbf{\bibinfo{volume}{117}}, \bibinfo{pages}{4511--4517} (\bibinfo{year}{2020}).

\bibitem{chaudhary2023geometric}
\bibinfo{author}{Chaudhary, G.}, \bibinfo{author}{Niu, L.}, \bibinfo{author}{Han, Q.}, \bibinfo{author}{Lewicka, M.} \& \bibinfo{author}{Mahadevan, L.}
\newblock \bibinfo{title}{Geometric mechanics of ordered and disordered kirigami}.
\newblock \emph{\bibinfo{journal}{Proceedings of the Royal Society A: Mathematical, Physical and Engineering Sciences}} \textbf{\bibinfo{volume}{479}} (\bibinfo{year}{2023}).

\bibitem{fulco2025disorder}
\bibinfo{author}{Fulco, S.}, \bibinfo{author}{Budzik, M.~K.}, \bibinfo{author}{Xiao, H.}, \bibinfo{author}{Durian, D.~J.} \& \bibinfo{author}{Turner, K.~T.}
\newblock \bibinfo{title}{Disorder enhances the fracture toughness of 2d mechanical metamaterials}.
\newblock \emph{\bibinfo{journal}{PNAS nexus}} \textbf{\bibinfo{volume}{4}}, \bibinfo{pages}{pgaf023} (\bibinfo{year}{2025}).

\bibitem{choukir2025disorder}
\bibinfo{author}{Choukir, S.}, \bibinfo{author}{Manohara, N.} \& \bibinfo{author}{Singh, C.~V.}
\newblock \bibinfo{title}{Disorder unlocks the strength-toughness trade-off in metamaterials}.
\newblock \emph{\bibinfo{journal}{Applied Materials Today}} \textbf{\bibinfo{volume}{42}}, \bibinfo{pages}{102579} (\bibinfo{year}{2025}).

\bibitem{reid2018auxetic}
\bibinfo{author}{Reid, D.~R.} \emph{et~al.}
\newblock \bibinfo{title}{Auxetic metamaterials from disordered networks}.
\newblock \emph{\bibinfo{journal}{Proceedings of the National Academy of Sciences}} \textbf{\bibinfo{volume}{115}}, \bibinfo{pages}{E1384--E1390} (\bibinfo{year}{2018}).

\bibitem{bastek2023inverse}
\bibinfo{author}{Bastek, J.-H.} \& \bibinfo{author}{Kochmann, D.~M.}
\newblock \bibinfo{title}{Inverse design of nonlinear mechanical metamaterials via video denoising diffusion models}.
\newblock \emph{\bibinfo{journal}{Nature Machine Intelligence}} \textbf{\bibinfo{volume}{5}}, \bibinfo{pages}{1466--1475} (\bibinfo{year}{2023}).

\bibitem{ha2023rapid}
\bibinfo{author}{Ha, C.~S.} \emph{et~al.}
\newblock \bibinfo{title}{Rapid inverse design of metamaterials based on prescribed mechanical behavior through machine learning}.
\newblock \emph{\bibinfo{journal}{Nature Communications}} \textbf{\bibinfo{volume}{14}}, \bibinfo{pages}{5765} (\bibinfo{year}{2023}).

\bibitem{jin2023recent}
\bibinfo{author}{Jin, H.}, \bibinfo{author}{Zhang, E.} \& \bibinfo{author}{Espinosa, H.~D.}
\newblock \bibinfo{title}{Recent advances and applications of machine learning in experimental solid mechanics: A review}.
\newblock \emph{\bibinfo{journal}{Applied Mechanics Reviews}} \textbf{\bibinfo{volume}{75}}, \bibinfo{pages}{061001} (\bibinfo{year}{2023}).

\bibitem{zheng2023deep}
\bibinfo{author}{Zheng, X.}, \bibinfo{author}{Zhang, X.}, \bibinfo{author}{Chen, T.-T.} \& \bibinfo{author}{Watanabe, I.}
\newblock \bibinfo{title}{Deep learning in mechanical metamaterials: from prediction and generation to inverse design}.
\newblock \emph{\bibinfo{journal}{Advanced Materials}} \textbf{\bibinfo{volume}{35}}, \bibinfo{pages}{2302530} (\bibinfo{year}{2023}).

\bibitem{reiser2022graph}
\bibinfo{author}{Reiser, P.} \emph{et~al.}
\newblock \bibinfo{title}{Graph neural networks for materials science and chemistry}.
\newblock \emph{\bibinfo{journal}{Communications Materials}} \textbf{\bibinfo{volume}{3}}, \bibinfo{pages}{93} (\bibinfo{year}{2022}).

\bibitem{khemani2024review}
\bibinfo{author}{Khemani, B.}, \bibinfo{author}{Patil, S.}, \bibinfo{author}{Kotecha, K.} \& \bibinfo{author}{Tanwar, S.}
\newblock \bibinfo{title}{A review of graph neural networks: concepts, architectures, techniques, challenges, datasets, applications, and future directions}.
\newblock \emph{\bibinfo{journal}{Journal of Big Data}} \textbf{\bibinfo{volume}{11}}, \bibinfo{pages}{18} (\bibinfo{year}{2024}).

\bibitem{dai2021graph}
\bibinfo{author}{Dai, M.}, \bibinfo{author}{Demirel, M.~F.}, \bibinfo{author}{Liang, Y.} \& \bibinfo{author}{Hu, J.-M.}
\newblock \bibinfo{title}{Graph neural networks for an accurate and interpretable prediction of the properties of polycrystalline materials}.
\newblock \emph{\bibinfo{journal}{npj Computational Materials}} \textbf{\bibinfo{volume}{7}}, \bibinfo{pages}{103} (\bibinfo{year}{2021}).

\bibitem{zheng2023unifying}
\bibinfo{author}{Zheng, L.}, \bibinfo{author}{Karapiperis, K.}, \bibinfo{author}{Kumar, S.} \& \bibinfo{author}{Kochmann, D.~M.}
\newblock \bibinfo{title}{Unifying the design space and optimizing linear and nonlinear truss metamaterials by generative modeling}.
\newblock \emph{\bibinfo{journal}{Nature Communications}} \textbf{\bibinfo{volume}{14}}, \bibinfo{pages}{7563} (\bibinfo{year}{2023}).

\bibitem{frey2025multi}
\bibinfo{author}{Frey, R.}, \bibinfo{author}{Tucker, M.~R.}, \bibinfo{author}{Afrasiabi, M.} \& \bibinfo{author}{Bambach, M.}
\newblock \bibinfo{title}{Multi-objective design of multi-material truss lattices utilizing graph neural networks}.
\newblock \emph{\bibinfo{journal}{Scientific Reports}} \textbf{\bibinfo{volume}{15}}, \bibinfo{pages}{3187} (\bibinfo{year}{2025}).

\bibitem{maurizi2025designing}
\bibinfo{author}{Maurizi, M.} \emph{et~al.}
\newblock \bibinfo{title}{Designing metamaterials with programmable nonlinear responses and geometric constraints in graph space}.
\newblock \emph{\bibinfo{journal}{Nature Machine Intelligence}} \textbf{\bibinfo{volume}{7}}, \bibinfo{pages}{1023--1036} (\bibinfo{year}{2025}).

\bibitem{dold2023differentiable}
\bibinfo{author}{Dold, D.} \& \bibinfo{author}{van Egmond, D.~A.}
\newblock \bibinfo{title}{Differentiable graph-structured models for inverse design of lattice materials}.
\newblock \emph{\bibinfo{journal}{Cell Reports Physical Science}} \textbf{\bibinfo{volume}{4}} (\bibinfo{year}{2023}).

\bibitem{felsch2024generative}
\bibinfo{author}{Felsch, G.} \& \bibinfo{author}{Slesarenko, V.}
\newblock \bibinfo{title}{Generative models struggle with kirigami metamaterials}.
\newblock \emph{\bibinfo{journal}{Scientific Reports}} \textbf{\bibinfo{volume}{14}}, \bibinfo{pages}{19397} (\bibinfo{year}{2024}).

\bibitem{yu2025expansion}
\bibinfo{author}{Yu, H.} \emph{et~al.}
\newblock \bibinfo{title}{Expansion limits of meshed split-thickness skin grafts}.
\newblock \emph{\bibinfo{journal}{Acta biomaterialia}} \textbf{\bibinfo{volume}{191}}, \bibinfo{pages}{325--335} (\bibinfo{year}{2025}).

\bibitem{genin2009functional}
\bibinfo{author}{Genin, G.~M.} \emph{et~al.}
\newblock \bibinfo{title}{Functional grading of mineral and collagen in the attachment of tendon to bone}.
\newblock \emph{\bibinfo{journal}{Biophysical journal}} \textbf{\bibinfo{volume}{97}}, \bibinfo{pages}{976--985} (\bibinfo{year}{2009}).

\bibitem{cs2025molecular}
\bibinfo{author}{CS~de Alcantara, A.} \emph{et~al.}
\newblock \bibinfo{title}{Molecular-scale interactions at mineralized collagen interfaces prevent network percolation, preserving compliance}.
\newblock \emph{\bibinfo{journal}{ACS nano}} \textbf{\bibinfo{volume}{19}}, \bibinfo{pages}{31350--31362} (\bibinfo{year}{2025}).

\bibitem{hu2015stochastic}
\bibinfo{author}{Hu, Y.} \emph{et~al.}
\newblock \bibinfo{title}{Stochastic interdigitation as a toughening mechanism at the interface between tendon and bone}.
\newblock \emph{\bibinfo{journal}{Biophysical Journal}} \textbf{\bibinfo{volume}{108}}, \bibinfo{pages}{431--437} (\bibinfo{year}{2015}).

\bibitem{olvera2020electroconductive}
\bibinfo{author}{Olvera, D.}, \bibinfo{author}{Sohrabi~Molina, M.}, \bibinfo{author}{Hendy, G.} \& \bibinfo{author}{Monaghan, M.~G.}
\newblock \bibinfo{title}{Electroconductive melt electrowritten patches matching the mechanical anisotropy of human myocardium}.
\newblock \emph{\bibinfo{journal}{Advanced Functional Materials}} \textbf{\bibinfo{volume}{30}}, \bibinfo{pages}{1909880} (\bibinfo{year}{2020}).

\bibitem{rogers2010materials}
\bibinfo{author}{Rogers, J.~A.}, \bibinfo{author}{Someya, T.} \& \bibinfo{author}{Huang, Y.}
\newblock \bibinfo{title}{Materials and mechanics for stretchable electronics}.
\newblock \emph{\bibinfo{journal}{science}} \textbf{\bibinfo{volume}{327}}, \bibinfo{pages}{1603--1607} (\bibinfo{year}{2010}).

\bibitem{xue2020mechanically}
\bibinfo{author}{Xue, Z.}, \bibinfo{author}{Song, H.}, \bibinfo{author}{Rogers, J.~A.}, \bibinfo{author}{Zhang, Y.} \& \bibinfo{author}{Huang, Y.}
\newblock \bibinfo{title}{Mechanically-guided structural designs in stretchable inorganic electronics}.
\newblock \emph{\bibinfo{journal}{Advanced Materials}} \textbf{\bibinfo{volume}{32}}, \bibinfo{pages}{1902254} (\bibinfo{year}{2020}).

\end{thebibliography}

\end{document}